\documentclass[5p,number,sort&compress]{elsarticle}

\usepackage{amsmath}

\journal{Physics Letters B}

\begin{document} 

\begin{frontmatter}

\title{Test of the Porter-Thomas distribution with the total cross section autocorrelation function}
 
\author[ncsu,tunl]{E. D. Davis}
\ead{dedavis4@ncsu.edu}

\address[ncsu]{Department of Physics, North Carolina State University, Raleigh, 
                         North Carolina 27695-8202, USA}
\address[tunl]{Triangle Universities Nuclear Laboratory, Durham, North Carolina 27708-0308, USA}

\begin{abstract} 
On the basis of experiments conducted in the 1970s and earlier, it is believed that, 
in a compound nucleus (CN), fluctuations in transition strengths are drawn from a 
Porter-Thomas distribution (PTD), a $\chi^2$ distribution of one degree of freedom. 
However, work done post-1990 at the Los Alamos Neutron Science Center 
(LANSCE) and Oak Ridge Electron Linear Accelerator (ORELA) facilities has yielded 
instances of neutron resonance data sets of superior quality that are inconsistent 
with the PTD.  In view of the importance of the status 
of the PTD to the foundations of statistical reaction theory, the question arises 
whether other data acquired at these facilities can be mined for evidence of 
deviations from the PTD?
To date, the focus has been on data taken in the regime 
of isolated resonances. In this letter, arguments are put forward in support of the 
analysis of measurements of the total cross section in the regime where resonances
are weakly overlapping.
\end{abstract} 
 
 
\end{frontmatter}
 
The Porter-Thomas distribution~\cite{PT56} for fluctuations in decay mode strengths of excited 
nuclei is an integral feature of conventional statistical models of the compound nucleus~\cite{MRW10}. 
Nevertheless, experiments conducted at LANSCE and ORELA in the last two decades have 
identified resonance data sets that are almost certainly statistically inconsistent with the
PTD~\cite{KBK10,KRU12,KBK13,KLG13}. Moreover, subsequent analyses~\cite{Ko11,SWM14} 
of the reduced neutron widths in the nuclear data ensemble seem to overturn the long
held belief that it furnishes persuasive evidence for the applicability of the Gaussian orthogonal 
ensemble (GOE) of Hamiltonian matrices~\cite{HPB82} in the description of CN fluctuation 
properties. There have been several attempts to reconcile these new findings with the standard
statistical models of CN
processes~\cite{We10,KBK11,CAI11,Vo11,SZ12,Mu15,FS15,VWZ15,MA16,Bo17,So17,We17,FBA18},
but there is a consensus~\cite{Re10,We14,Au15} that more data is needed to guide theoretical 
considerations.

The analyses of the experiments cited above have been confined to the resolved resonance 
regime. However, the measurements performed extend into the unresolved resonance regime, 
and invariably include data on total cross sections~\cite{KG13}. This archived data could be used
to test statistical reaction models via appropriate correlation functions.
The suggested tool for the analysis of the data on the total cross section $\sigma_\text{tot}(E)$
is the autocorrelation function 
\begin{equation} \label{eq:Rdef}
 R_\text{tot}(\varepsilon) = \frac{\left\langle\sigma_\text{tot}(E+\tfrac{1}{2}\varepsilon)
                                \sigma_\text{tot}(E-\tfrac{1}{2}\varepsilon)\right\rangle}{
                                 \left\langle\sigma_\text{tot}\right\rangle^2} - 1 ,
\end{equation}
where the angle brackets denote averages over the scattering energy. [Implicit in 
Eq.~(\ref{eq:Rdef}) is the assumption of stationarity of the averages, i.e., they are 
independent of the energy $E$.] With a few exceptions (viz., Ref.~\cite{BBD18}), 
autocorrelation function studies of nuclear reactions have previously been confined 
to the regime of strongly overlapping resonances~\cite{MRW10}. In the present work, it is 
advocated that $R_\text{tot}(\varepsilon)$ be determined in the weakly overlapping 
resonance regime.   

Why consider a correlation function involving the \emph{total\/} cross section as opposed to 
any other type of cross section?
Via the optical theorem, the fluctuations probed by $R_\text{tot}(\varepsilon)$ can be 
related to the two-point measure
\begin{equation} \label{eq:C2ab}
 C_{ab}(\varepsilon ) = \left\langle S_{aa}^{\text{fl}*}(E+\tfrac{1}{2}\varepsilon)  
                                                    S_{bb}^{\text{fl}}(E-\tfrac{1}{2}\varepsilon)  \right\rangle
\end{equation}
of fluctuations in elastic elements of the $S$-matrix ($S^\text{fl}\equiv S-\left\langle 
S \right\rangle$). Of significance in the present context is that the contributions to 
$R_\text{tot}(\varepsilon)$ for which the channels $a$ and $b$ in $C_{ab}(\varepsilon)$ 
coincide 
involve the \emph{variance\/} of partial widths for the entrance channel. Thus, there is
a direct link between $R_\text{tot}(\varepsilon)$ for neutron-induced reactions and fluctuations in 
reduced neutron widths. This aspect of $R_\text{tot}(\varepsilon)$ was recognized by Ericson 
in his seminal treatment of cross section fluctuations~\cite{Er63}.

Another important feature of $R_\text{tot}(\varepsilon)$ concerns its computation within models 
of compound nucleus reactions for medium-weight and heavy nuclei. At present, the 
gold standard for such models is a stochastic treatment of resonance reactions invoking 
the GOE of Hamiltonian matrices, which is outlined in Sec.~IV.B of Ref.~\cite{MRW10} and 
is hereafter referred to as the Heidelberg model. Since the ground-breaking work of 
Verbaarschot~\emph{et al.}~\cite{VWZ84,VWZ85} on the exact calculation of averages 
within the Heidelberg model in the limit of infinitely many resonances, it has been known that
$C_{ab}(\varepsilon )$ and, hence, $R_\text{tot}(\varepsilon)$ can be expressed 
as three dimensional integrals. Not only do these integral representations hold for any 
number of open channels and in all resonance regimes (from isolated to strongly overlapping), 
the input required for their evaluation is limited to the average $S$-matrix itself and, when
$\varepsilon\not=0$, the average level spacing (for each spin). Numerical evaluation of 
the integrals is demanding but feasible, and has been carried out in a number of comparisons 
of various statistical approaches to low-energy compound nucleus 
reactions~\cite{Ve86,DB88,HLK03,DHR10,KTW15}.

In principle, then, parameter-free predictions for $R_\text{tot}(\varepsilon)$ are possible in 
the Heidelberg model, and any large differences between these predictions and data 
cannot be attributed to approximations in the evaluation of $R_\text{tot}(\varepsilon)$ 
within the model. It is significant that the quite different maximum entropy approach to 
statistical nuclear reactions~\cite{MPS85,FM85} is known to yield results for averages 
like those in $R_\text{tot}(0)$ which are in complete agreement with the corresponding results 
obtained within the Heidelberg model when the number of resonances is infinite~\cite{Br95}. 
Use of the Heidelberg model is tantamount to the adoption of the most probable unbiased distribution 
of $S$-matrix fluctuations consistent with unitarity and causality.  
Non-generic dynamical effects are not accommodated.

What is the potential size of any discrepancy between empirical values of
$R_\text{tot}(\varepsilon)$ and theoretical estimates deduced from the Heidelberg model? 
To address this issue, it is advantageous to consider the 
approximation of $C_{ab}(\varepsilon )$ in the statistical Breit-Wigner (SBW) model using the 
scheme of calculation laid out in Ref.~\cite{EDR16}. Recent work along these lines in the 
analysis of fluctuations in the ${}^{235}$U fission cross section has shown that the SBW 
model can yield results which account quantitatively for the features in the isolated  
resonance regime of the cross section autocorrelation function studied~\cite{BBD18}.

In the SBW model, it is possible to relax the assumption that partial widths are drawn 
from the PTD. Instead, guided by the empirical characterisation of data on partial widths in 
Ref.~\cite{KBK10}, it is assumed that partial widths are drawn from a $\chi^2$ distribution 
of $\nu$ degrees of freedom, In the weakly overlapping resonance regime, it is found that 
the dominant contribution to $C_{ab}(0)$ is
\begin{equation}  \label{eq:Cns}
 C_{ab}^{(d)}(0) = \left( 1 + \frac{2}{\nu}\delta_{ab} \right) T_a T_b \, I_{ab}^{(\nu)} ,
\end{equation}
where the transmission coefficients $T_c = 1 - \left| \left\langle S_{cc} \right\rangle \right|^2$, 
and
\begin{equation} 
 I_{ab}^{(\nu)} = \int\limits_0^\infty  \frac{\prod_c \left( 1+\tfrac{2}{\nu}T_c\tau \right)^{-\nu/2}}{
                                 \left(1+\tfrac{2}{\nu}T_a\tau\right)\left(1+\frac{2}{\nu}T_b\tau\right)} d\tau  .
\end{equation}
(The product in the numerator of the integrand above is over all open channels $c$.)

\begin{figure}[tb] 
\begin{center} 
\includegraphics[width=\columnwidth]{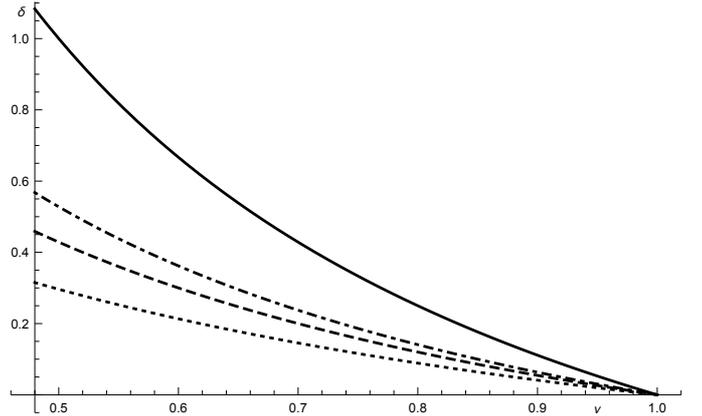}
\caption{The relative change  of the dominant contribution to $C_{aa}(0)$ in the SBW 
mode [i.e., $\delta$ in Eq.~(\ref{eq:ddef})] versus $\nu$ for different choices of the number 
$\Lambda$ of open channels: $\Lambda=5$ (dotted line), $\Lambda=10$ (dashed line), 
$\Lambda=20$ (dot-dashed line), and $\Lambda=\infty$ (solid line).}
\label{fg:nuDep} 
\end{center} 
\end{figure}

The $\nu$ dependence in Eq.~(\ref{eq:Cns}) is encouraging. Figure \ref{fg:nuDep} displays 
the relative change 
\begin{equation} \label{eq:ddef}
  \delta \equiv \frac{C^{(d)}_{aa}(0)}{C^{(d)}_{aa}(0)\ [\nu=1]}  - 1
\end{equation}
in the dominant contribution to $C_{aa}(0)$ as $\nu$ ranges from its value in the Porter-Thomas 
limit ($\nu=1$) through values implied by the analysis of Pt neutron width data 
($\nu\approx\tfrac{1}{2}$). 
[In Eq.~(\ref{eq:ddef}), the value of the 
dominant contribution to $C_{aa}(0)$ for arbitrary $\nu$ is divided by its value for $\nu=1$.] 
In generating Fig.~\ref{fg:nuDep}, all transmission coefficients have, for simplicity, been
taken to be equal in all $\Lambda$ open channels, 
which means that $\delta$ is a function of only $\nu$ and $\Lambda$. 

For values of $\nu$ comparable to those found in the statistical analysis of reduced neutron
widths in Ref.~\cite{KBK10},
Fig.~\ref{fg:nuDep} suggests that $R_\text{tot}(0)$ could deviate from its value in the Heidelberg 
model by more than 20\%. Even allowing for uncertainties in the transmission coefficients 
needed to evaluate the theoretical expression for $R_\text{tot}(0)$ within the Heidelberg model, 
this should be a large enough 
signal to warrant determination of $R_\text{tot}(0)$ with high quality total cross section data for 
weakly overlapping resonances in the unresolved resonance regime.

This letter has discussed a test of the Heidelberg model involving fluctuations in 
neutron-induced CN reactions. For weakly overlapping resonances, $R_\text{tot}(0)$ is sensitive
to fluctuations in reduced neutron widths but insensitive to correlations between levels (beyond
level repulsion). These properties make the study of $R_\text{tot}(0)$ for weakly overlapping 
resonances in the unresolved resonance regime a test, in effect, of the Porter-Thomas distribution. 
To date, there have been
no investigations of this kind, but chaotic two-dimensional microwave resonator data
has been used to test the Verbaarschot \emph{et al.\/}~result for $C_{ab}(\varepsilon )$ in the 
weakly overlapping resonance regime, and good agreement was found~\cite{DFH08}. 
More recently, in a tour de force (building on the supersymmetric methodology of 
Refs.~\cite{KNS13,NKS14}), Kumar \emph{et al.\/}~have managed to derive, within the
framework of the Heidelberg model, a four dimensional integral representation for the characteristic 
function of $\sigma_{ab}(E)=\left|S^\text{fl}_{ab}(E)\right|^2$ when $a\not=b$~\cite{KDG17}, which,
like the result of Verbaarschot \emph{et al.\/}~for $C_{ab}(\epsilon)$, is exact in the limit of infinitely 
many resonances and holds for any number of open channels and in all resonance regimes. 
This result for the characteristic function was used in a comparison with an experimental 
cross distribution inferred from weakly overlapping resonance data \cite{CAP60} for the reaction 
${}^{37}\text{Cl}(p,\alpha){}^{34}\text{S}$. As observed in Ref.~\cite{KDG17},
it is possible to conclude that the characteristic function is more than capable of reproducing the 
experimental cross distribution but there are 
two shortcomings to the comparison. First, the quality of the data could be significantly improved 
upon; only 51 of an estimated 120 or so resonances were observed in the energy interval of interest,
and, more worryingly, no attempt was made to identify the resonance spins: the disturbing 
conclusions~\cite{KBK10,Ko11,KBK13} 
drawn from reduced neutron width data rest on the careful identification of resonances of a given 
spin and parity. The other deficiency in the comparison is the fact that adjustments to
the number of effective channels  and the associated transmission coefficient
are made to obtain agreement. For a test of the kind contemplated in the present paper,
a comparison free of fit parameters should be performed.

\section*{Acknowledgements}

This work was supported in part by the US Department of Energy under Grant 
No.~DE-FG02-97ER41042. 
I would like to thank Dr. P. E. Koehler for his interest in this work.

\section*{References}

\bibliography{cfptdbibfile}

\end{document}